\documentclass[
     reprint,
     superscriptaddress,
     amsmath,amssymb,amsfonts,
     aps,
     prmaterials,
     floatfix,
    ]{revtex4-2}

\usepackage{xfrac}
\usepackage{braket}
\usepackage{graphicx} \graphicspath{{PICS/}}
\usepackage{verbatim}
\usepackage{etoolbox}
\usepackage[utf8]{inputenc}
\usepackage{mathtools}
\usepackage{soul,xcolor,colortbl}
\usepackage{hyperref} \hypersetup{colorlinks=true,citecolor=blue,linkcolor=blue,urlcolor=blue}
\usepackage{tabularx}
\usepackage{csquotes}
\usepackage{multirow}
\usepackage{longtable}
\usepackage{cellspace}
\usepackage{resizegather}
\usepackage{dcolumn}
\usepackage[normalem]{ulem}
\usepackage{siunitx}
\DeclareSIUnit\angstrom{\text{\AA}}
\usepackage{physics}
\usepackage{pdfpages}
\usepackage{placeins}
\usepackage{orcidlink}

\usepackage{bm}
\usepackage{array}
\newcolumntype{C}[1]{>{\centering\arraybackslash}p{#1}}

\usepackage{color}
\usepackage{morefloats}

\definecolor{Gray}{gray}{0.9}
\definecolor{LightCyan}{rgb}{0.88,1,1}
\definecolor{green}{rgb}{0.5451,0.2706,0.0745}
\definecolor{changed1}{rgb}{0,0,0}
\definecolor{changed2}{rgb}{0,0,0}
\definecolor{changed3}{rgb}{0,0,1}
\definecolor{changed4}{rgb}{1,0,0}

\makeatletter
\AtBeginDocument{\let\LS@rot\@undefined}
\makeatother

\begin{document}

\title{Kagome Flat Bands from Self-Assembled Water on Non-van der Waals 2D Materials}

\author{Tom Barnowsky\,\orcidlink{0000-0003-1626-4644}}
\affiliation{Theoretical Chemistry, Technische Universit\"at Dresden, 01062 Dresden, Germany}
\affiliation{Institute of Ion Beam Physics and Materials Research, Helmholtz-Zentrum Dresden-Rossendorf, 01328 Dresden, Germany}
\author{Madeleine Christie\,\orcidlink{0009-0007-8818-9678}}
\thanks{Present address: Centre for Astrophysics Research, University of Hertfordshire, Hatfield AL10 9AB, United Kingdom}
\affiliation{Institute of Ion Beam Physics and Materials Research, Helmholtz-Zentrum Dresden-Rossendorf, 01328 Dresden, Germany}
\author{Anastasiia Nihei\,\orcidlink{0009-0006-0851-709X}}
\affiliation{Theoretical Chemistry, Technische Universit\"at Dresden, 01062 Dresden, Germany}
\affiliation{Institute of Ion Beam Physics and Materials Research, Helmholtz-Zentrum Dresden-Rossendorf, 01328 Dresden, Germany}
\author{Rico Friedrich\,\orcidlink{0000-0002-4066-3840}}
\email[]{rico.friedrich@tu-dresden.de}
\affiliation{Theoretical Chemistry, Technische Universit\"at Dresden, 01062 Dresden, Germany}
\affiliation{Institute of Ion Beam Physics and Materials Research, Helmholtz-Zentrum Dresden-Rossendorf, 01328 Dresden, Germany}

\date{\today}

\begin{abstract}
The properties of two-dimensional (2D) materials are governed not only by their intrinsic electronic structure but also in particular by their surface chemistry and interactions with the environment.
The surfaces of non-van der Waals (non-vdW) 2D materials are intrinsically reactive and may require chemical stabilization under realistic experimental conditions.
Here, we perform a systematic first-principles study of water passivation of non-vdW 2D materials and demonstrate that molecular water adsorption is energetically and dynamically favored for a wide range of systems.
Rather than dissociating, water molecules frequently remain intact and form adsorbed ordered surface overlayers.
These overlayers {can} adopt a kagome lattice geometry, leading to the emergence of flat electronic bands {directly} at the valence-band edge.
{Analysis of the electronic structure reveals a {strong intermolecular coupling between neighboring water lone pairs within this kagome topology}, while the locked molecular orientations simultaneously produce an electrostatically frustrated surface potential.}
These findings establish water not only as a stabilizing agent but also as an active element for electronic structure engineering of non-vdW 2D surfaces.

\vspace{0.2cm}

\end{abstract}

\maketitle

\section{Introduction} \label{intro}
Two-dimensional (2D) materials exhibit electronic, magnetic, and chemical properties distinct from their bulk counterparts, making them attractive for applications ranging from electronics~\cite{Lemme_NCOM_2022} to catalysis~\cite{Deng_NNano_2016}.
While research has largely focused on van der Waals (vdW) layered compounds, there is growing interest in the new class of non-vdW 2D materials {extracted from non-layered crystals}~\cite{Puthirath_Balan_MatTod_2022,Kaur_AdvMat_2022,Barnowsky_NatCommun_2026}.

{However, unlike the weak interlayer forces present in vdW materials, non-vdW compounds are held together by chemical bonds; cleavage of these bonds leaves surface atoms undercoordinated, exposing dangling bonds and rendering non-vdW 2D sheets highly reactive -- often leading to geometric reconstructions, surface states, or even environmental degradation~\cite{Kaur_AdvMat_2022}.}
Stabilizing such surfaces while enabling enhancement of their properties remains a major challenge, limiting both experimental realization and functional exploitation of non-vdW 2D systems~\cite{Gao_ACSNano_2024}.

In experiments, these materials are commonly produced via liquid-phase exfoliation (LPE)~\cite{Nicolosi_Science_2013,Puthirath_Balan_NNANO_2018}.
Even when nominally water-free solvents are employed, trace amounts of water are difficult to eliminate completely.
In addition, many applications require environmental stability under ambient conditions.
As a result, interactions between non-vdW 2D surfaces and water molecules are often unavoidable and can play a decisive role in determining surface stability and electronic properties~\cite{Wei_JPCC_2020}.
However, the impact of molecular water adsorption on the electronic structure of non-vdW 2D materials remains largely unexplored.

From an interface-design perspective, controlling such surface interactions is therefore essential.
{Surface passivation is typically viewed primarily as a means to suppress dangling bonds and stabilize the surface.
In contrast, a recent study on non-vdW 2D systems has demonstrated that passivating species can themselves introduce new functionalities, beyond mere stabilization~\cite{Barnowsky_NanoLett_2024}.}
In this work, we show that water on non-vdW 2D surfaces not only stabilizes the materials but can also self-assemble into an ordered kagome lattice, giving rise to flat electronic bands near the valence-band edge with a singular density of states (DOS).

Unlike moir\'e superlattices~\cite{Cao_Nature_2018,Debnath_ACSAMI_2025} or lithographically defined patterns~\cite{Weinhold_RSCAdv_2021}, this approach relies on spontaneous molecular ordering at the surface, offering a chemically intuitive and potentially scalable pathway to flat-band engineering.
This reveals water passivation as a versatile route to simultaneously stabilize and engineer the electronic properties of non-vdW 2D materials.

\section{Results and Discussion}

\begin{figure*}[ht!]
    \centering
    \includegraphics[width=.9\linewidth]{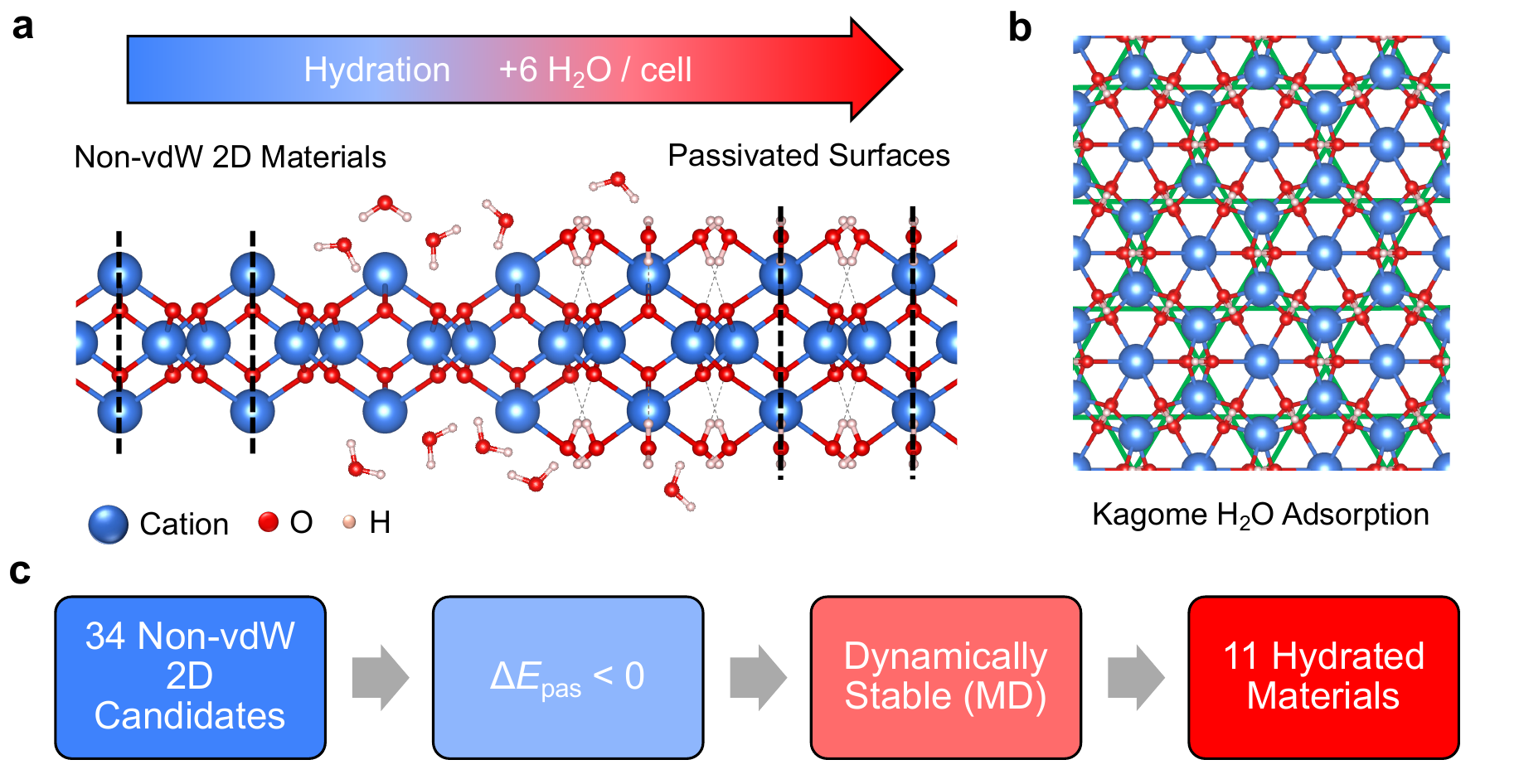}
    \caption{\small\textbf{Water Passivation of Non-vdW 2D Materials.}
    (\textbf{a}) Schematic illustration of the hydration of a non-vdW 2D material, showing adsorption of six H$_2$O molecules per unit cell {(highlighted by black dashed lines)} at the exposed surface sites created upon exfoliation of the 2D system from {the bulk parent system}.
    Water molecules bind to surface cations through oxygen, while one hydrogen atom orients towards {an anion of the slab}, resulting in complete passivation of dangling bonds.
    (\textbf{b}) Top view highlighting the adsorption sites of the water molecules, forming a kagome lattice on the surface (highlighted by green lines).
    (\textbf{c}) Overview of the {high-throughput} screening workflow.
    Non-vdW 2D candidates are filtered based on negative passivation energies and dynamic stability, yielding a subset of hydrated 2D materials.
    }
    \label{fig1}
\end{figure*}

To assess whether water can act as a viable and thermodynamically favorable passivant for non-vdW 2D materials, we systematically {investigate} water-surface interactions across a broad set of 34 previously predicted non-vdW 2D compounds~\cite{Friedrich_NanoLett_2022,Barnowsky_AdvElMats_2023}.
Our analysis first focuses on the energetic driving force for water adsorption relative to the pristine {material} and free water molecules, as well as the stability of the resulting surface-bound configurations.

For this purpose, we define a passivation energy $\Delta E_\mathrm{pas}$, which quantifies the energetic preference for hydration compared to the isolated reference systems.
As outlined in Fig.~\ref{fig1}\textbf{a}, dense passivation of all dangling bonds leads to six H$_2$O molecules per unit cell.
The passivation energy is hence defined as
\begin{equation} \Delta E_\mathrm{pas} = \frac{1}{6} \qty(E_\mathrm{passivated}-E_\mathrm{pristine}-6E_\mathrm{H_2O})
\end{equation}
following Ref.~\cite{Barnowsky_NanoLett_2024}.
Negative values of $\Delta E_\mathrm{pas}$ indicate that water adsorption is thermodynamically preferred.
A substantial fraction -- 31 out of the investigated 34 systems -- exhibit negative passivation energies, suggesting that interaction with water is generally favorable in contrast to hydrogenation~\cite{Barnowsky_NanoLett_2024}.
Remarkably, for a large fraction of systems ($23/34 \approx 68\%$), the water molecules remain intact upon adsorption (rather than dissociating) and self-assemble into an ordered kagome surface lattice as shown in Fig.~\ref{fig1}\textbf{b}.

Thermodynamic favorability alone, however, does not guarantee structural stability~\cite{Malyi_ACSAMI_2019}.
As a high-throughput compatible validation, we further assess the dynamical stability of the water-passivated {materials} at \SI{300}{\kelvin}.
{Only systems whose energy remains stable during \SI{5}{\pico\second} finite-temperature molecular dynamics (MD) simulations are considered dynamically stable and are retained for subsequent structural and electronic analysis.}
This completes the utilized high-throughput pipeline to find thermodynamically preferred, yet stable hydrated non-vdW 2D materials, outlined in Fig.~\ref{fig1}\textbf{c}.

\begin{figure*}[ht!]
    \centering
    \includegraphics[width=\linewidth]{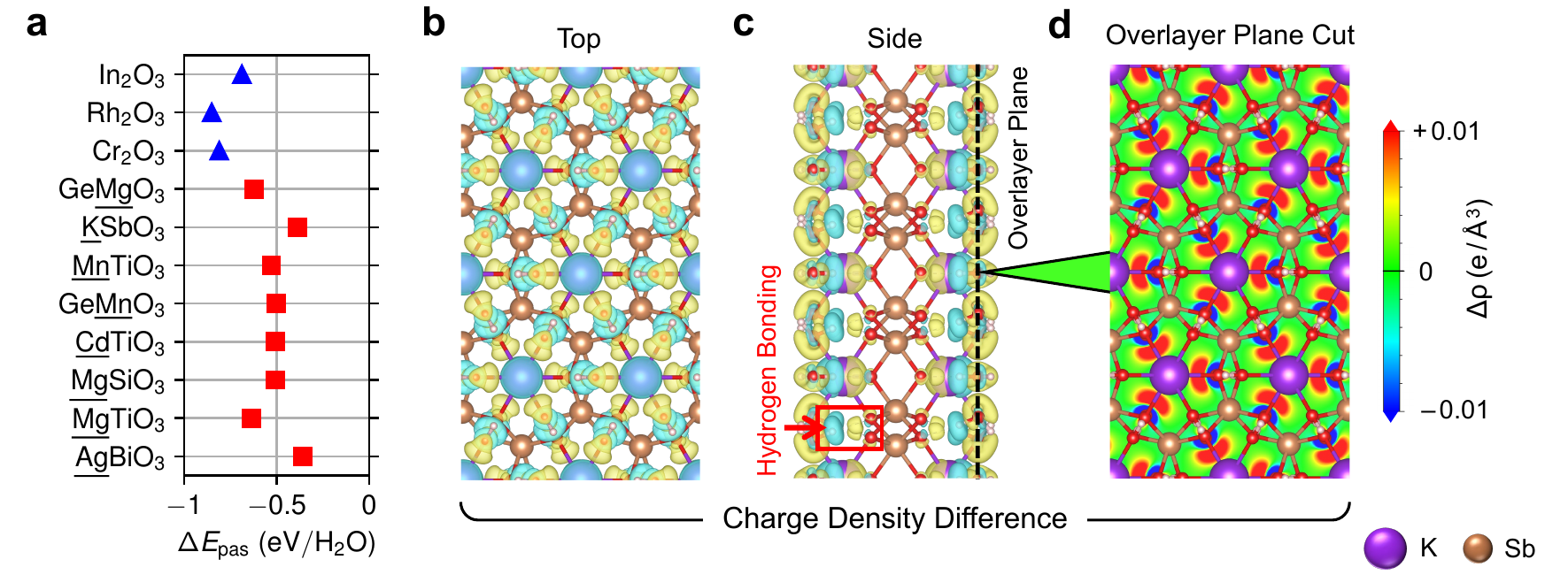}
    \caption{\small\textbf{Thermodynamic Properties and Charge Density Difference.}
    (\textbf{a}) Passivation energies for the final set of thermodynamically favored and dynamically stable non-vdW 2D materials upon hydration.
    Systems for which water dissociates upon bonding are marked with blue triangles, while systems where water molecules remain intact are shown as red squares.
    {For ternary compounds, the surface terminating species is underlined in the chemical formula.}
    (\textbf{b} \& \textbf{c}) Top and side view of the charge-density difference upon H$_2$O-passivation of 2D KSbO$_3$.
    Yellow and blue isosurfaces correspond to charge accumulation and depletion, respectively (isosurface level  \SI{0.01}{e \per \angstrom^3}); {K appears blue due to the enveloping isosurface.}
    The newly formed hydrogen bonds are indicated for an example region using a red box.
    (\textbf{d}) Planar cut through the charge density difference of the top overlayer.
    The position of the cut is indicated in panel (\textbf{c}) using a dashed line.
    The bottom overlayer exhibits an identical pattern rotated by \SI{60}{\degree} due to the hexagonal symmetry.
    }
    \label{fig2}
\end{figure*}

Out of 34 initial structures, 11 remain after the MD filtering step.
Their passivation energies are {plotted} in Fig.~\ref{fig2}\textbf{a}.
Typical bonding energies lie in the range of several hundred meV per H$_2$O.
The interaction is distinctly stronger than in the case of hydrogenation, where passivation energies only surpass \SI{-0.5}{\eV\per H_2} in a single case out of 35 investigated 2D systems~\cite{Barnowsky_NanoLett_2024}.
The energetic penalty associated with the loss of entropy {$(-T\Delta S)$} upon passivation can also be estimated.
{
As an upper bound, one may assume that the entropy of free water molecules is entirely lost upon adsorption.
Using the tabulated room-temperature (\SI{300}{\kelvin}) entropy of gaseous water, \SI{188.8}{\joule\per\kelvin\per\mole}~\cite{Chase_NIST_JANAF_thermochem_tables_1998}, {this amounts to a $(-T\Delta S)$ of} approximately \SI[retain-explicit-plus]{+0.59}{\eV\per H_2O}.
Under this upper-bound estimate, the entropic cost exceeds the computed passivation energy for several of the more weakly bound systems in Fig.~\ref{fig2}\textbf{a}.
{Hence,} room-temperature stability is not guaranteed in these cases, while the more strongly bound systems remain favorable even under this conservative estimate.
As this penalty scales linearly with temperature and vanishes for $T\to0$, it is reduced at lower temperature, so {that} the weakly bound systems are expected to become thermodynamically stable upon cooling.
}

For these 11 dynamically stable passivated systems, two distinct structural patterns are observed:
(i) Dissociated water, \emph{i.e.}, OH$^-$ at the surface cations and H$^+$ at the anion sites of the slab {and} (ii) H$_2$O molecules bound to the surface cations via the oxygen atom.
Type (i) dissociated passivation is favored for the binary hematene type structures~\cite{Puthirath_Balan_NNANO_2018} while intact water molecules are observed in the case of the ternary ilmenene-like structures~\cite{Puthirath_Balan_CoM_2018}.

For the type (ii) example KSbO$_3$, Figs.~\ref{fig2}\textbf{b}--\textbf{d} show the charge-density difference of the passivated slab and its constituents, defined as
\begin{equation}
    \Delta \rho = \rho_\mathrm{passivated}-\rho^\mathrm{static}_\mathrm{pristine}-\rho^\mathrm{static}_\mathrm{H_2O}.
\end{equation}
Here, $\rho_\mathrm{passivated}$, $\rho^\mathrm{static}_\mathrm{pristine}$, and $\rho^\mathrm{static}_\mathrm{H_2O}$ denote the electron densities of the passivated {material}, the bare {sheet}, and the isolated H$_2$O molecules, respectively. The densities $\rho^\mathrm{static}_\mathrm{pristine}$ and $\rho^\mathrm{static}_\mathrm{H_2O}$ {are} computed for the same geometries as in the passivated configuration.
This construction uncovers charge rearrangements induced by adsorption and bonding. The pronounced redistribution indicates that the water molecules are strongly bound to the surface rather than being physisorbed.
The charge-density difference suggests a dominant interaction between the oxygen atom of H$_2$O and surface cations, consistent with a coordination-type bonding motif.
In addition, a weaker charge redistribution is observed between one hydrogen atom of each water molecule and {the} oxygen atoms of the slab, indicative of hydrogen bonding.
Although weak, this secondary interaction introduces a subtle coupling that perturbs the rotational freedom of the molecules, effectively locking them into a preferred orientation with respect to the surface.
{
Based on the analysis of {Bader} charges~\cite{bader1,bader2,bader3,bader4} in bulk, 2D, and hydrated 2D KSbO$_3$, the oxidation state of the surface ions does not change upon exfoliation or water passivation~\cite{Friedrich_CCE_2019}.
}

\begin{figure*}[ht!]
    \centering
    \includegraphics[width=\linewidth]{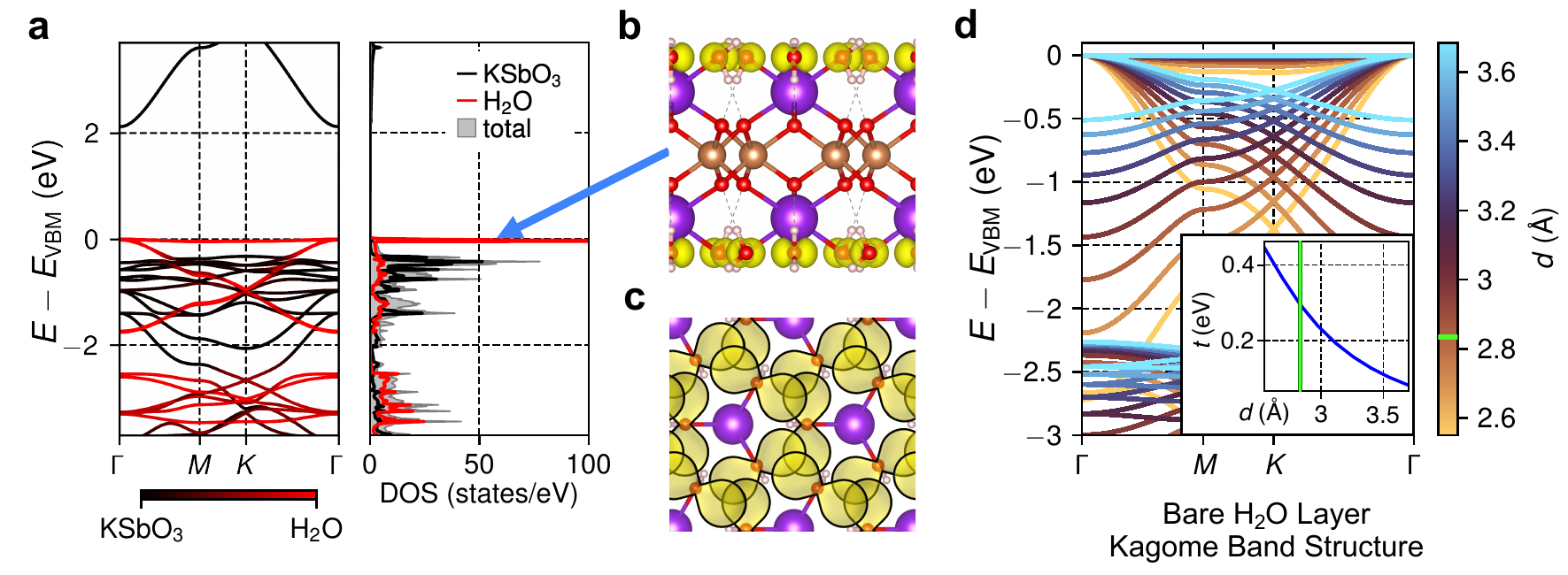}
    \caption{\small\textbf{Electronic Properties and Surface Electrostatics.}
    (\textbf{a}) Projected electronic band structure and corresponding projected DOS of H$_2$O-passivated 2D KSbO$_3$.
    Water-derived states including the flat band at the valence-band maximum are indicated in red, while the electronic states of the underlying 2D slab are shown in black.
    (\textbf{b}) Isosurface of the real-space partial charge density distribution associated with the flat valence band in hydrated KSbO$_3$, illustrating its strong localization on the molecular overlayer (isosurface level \SI{0.1}{e \per \angstrom^3}).
    {
    (\textbf{c}) Top view of the water molecules and K ions at the top surface including a schematic depiction of orientation and overlap of the water {lone pair} HOMOs.
    (\textbf{d}) Electronic band structure of the isolated H$_2$O overlayer.
    Different colors correspond to different intermolecular distances $d$ between neighboring water molecules.
    The inset shows the effective nearest-neighbor tight-binding coupling $t$, extracted from the energy splitting of the dispersive kagome bands, as a function of $d$.
    The intermolecular distance corresponding to the geometry of the water molecules adsorbed on the KSbO$_3$ surface is highlighted in {green} in both the band structure color code and the inset.
    }
    }
    \label{fig3}
\end{figure*}

While both dissociated and intact water configurations effectively passivate dangling bonds and thereby stabilize non-vdW {materials}, they lead to markedly different electronic signatures.
The complete set of band structures is provided in Sec.~I of the Supporting Information (SI).
For dissociated adsorption -- potentially relevant for water-splitting applications~\cite{Jun_EnergyAdv_2023} -- there are naturally no well-defined water states anymore, and the electronic states of the OH$^-$ and H$^+$ groups fully hybridize with the slab.
In contrast, when water remains intact, its molecular orbitals are largely preserved and, within a specific energy window, collectively form a band signature dictated by the kagome lattice arrangement.

A kagome lattice is well known to host characteristic electronic features arising from its topology, most notably a nearly dispersionless flat band together with dispersive bands that intersect at a Dirac-like crossing.
The flat band originates from destructive interference of hopping paths on the corner-sharing triangular motifs of the lattice, leading to strong real-space localization of the corresponding electronic states~\cite{Sutherland_PRB_1986,Leykam_APX_2018}.
{In contrast, the dispersive bands reflect itinerant states delocalized across the lattice.}
Under additional symmetry breaking conditions, kagome lattices have also been outlined to host topologically nontrivial band structures~\cite{Li_SciAdv_2018,Checkelsky_NatRevMater2024}.

In Fig.~\ref{fig3}\textbf{a} this is clearly exemplified by a projected band structure of H$_2$O-passivated 2D KSbO$_3$.
The water states in this energy region form the known kagome motif, with the flat band acting as the valence band maximum (VBM) leading to a singularity in the DOS.
{As the slab {has} a water overlayer on {each} surface, this flat band is doubly degenerate, with one such band arising from each surface.}
Especially noteworthy is the energetic isolation of the flat band above the non-vdW 2D material's {host states}.
Even for the lower-energy dispersive kagome states, where there is energetic overlap, the states of the pristine 2D sheet are virtually unchanged by the passivation.

{
{As shown in Fig.~\ref{fig3}\textbf{b}, the states forming the flat band are derived from the highest occupied molecular orbitals (HOMOs) of water -- the $1b_1$ lone-pair orbitals, built primarily from an oxygen $2p$ orbital {oriented parallel to the surface}.}
These states exhibit negligible hybridization with the electronic states of the underlying KSbO$_3$ sheet.
Their orientation, visualized in Fig.~\ref{fig3}\textbf{c}, positions neighboring orbital lobes to face each other, enabling strong electronic coupling.
This contrasts with a conventional setup, such as a carbon kagome lattice{~\cite{Springer_ChemSocRev_2020}}, where the contributing orbitals typically face out of plane, further emphasizing the unique coupling mechanism in this system.
Additionally, the orbitals are aligned perpendicular to the K-O bond, minimizing overlap with the potassium states.
Consequently, the electronic coupling responsible for the kagome bands arises predominantly from intermolecular orbital overlap within the water overlayer, rather than {host-mediated} interactions.
This is further demonstrated in Fig.~\ref{fig3}\textbf{d}, where the characteristic kagome band structure persists even when considering the isolated water overlayer alone.
}

{
Within a simple nearest-neighbor tight-binding description of an ideal kagome lattice {with hopping amplitude $t$}, the two dispersive bands are separated at the $\Gamma$ point by an energy $W=6t$~\cite{Ohgushi_PRB_2000}.
In this approximation, the effective electronic coupling between neighboring water HOMOs can therefore be estimated as $t = W/6$, with $W$ denoting the bandwidth of the dispersive kagome bands.
Consistent with the picture of intermolecular HOMO overlap giving rise to the kagome-derived electronic structure, the coupling strength depends sensitively on the intermolecular distance $d$, as shown in the inset of Fig.~\ref{fig3}\textbf{d}.
For the present geometry, corresponding to a molecular center-to-center distance of $d=\SI{2.83}{\angstrom}$, we obtain a {strong} effective coupling of $t\approx\SI{0.3}{\eV}$, both for the isolated water overlayer and for the water-passivated KSbO$_3$ surface.
}

The destructive interference of the flat-band gives rise to highly localized states and hence a pronounced peak in the DOS at the VBM, {directly observable in experiment}, while the conduction bands and deeper valence bands of the slab remain essentially unchanged.
The coexistence of highly localized kagome states with the extended states of the 2D sheet effectively decouples the molecular flat-band physics from the host electronic structure.
{Specifically, angle-resolved photoemission spectroscopy (ARPES) should {be able to} resolve the flat band as a dispersionless, high-intensity feature at the valence-band edge across the Brillouin zone.
{It is thus} clearly distinguishable from the dispersive bands of {the underlying 2D material} further below in energy, as has been demonstrated for electronic kagome flat bands in other systems~\cite{Li_SciAdv_2018,Zhang_NatRev_2022_ARPES}.}
The high DOS associated with the flat band provides a potential reservoir of hole states at the VBM, which may influence carrier dynamics and surface-mediated electronic processes~\cite{Leykam_APX_2018}.
{Because the flat band coincides with the VBM, modest hole doping -- for instance via chemical or electrostatic gating -- would be sufficient to partially populate it, a prerequisite for accessing the correlated or topological phenomena typically associated with partially filled kagome flat bands~\cite{Checkelsky_NatRevMater2024}.}
These features highlight the unique ability of molecular water overlayers to introduce well-defined electronic states at the surface while preserving the overall character of the underlying 2D material.

{
{We note that at} finite temperatures, MD simulations indicate that thermal fluctuations in the molecular orientations can broaden the kagome-related features and reduce the intermolecular coupling.
This implies that the electronic signatures are most clearly resolved in highly ordered or symmetry-constrained configurations.
}

\begin{figure}[t!]
    \centering
    \includegraphics[width=.8\linewidth]{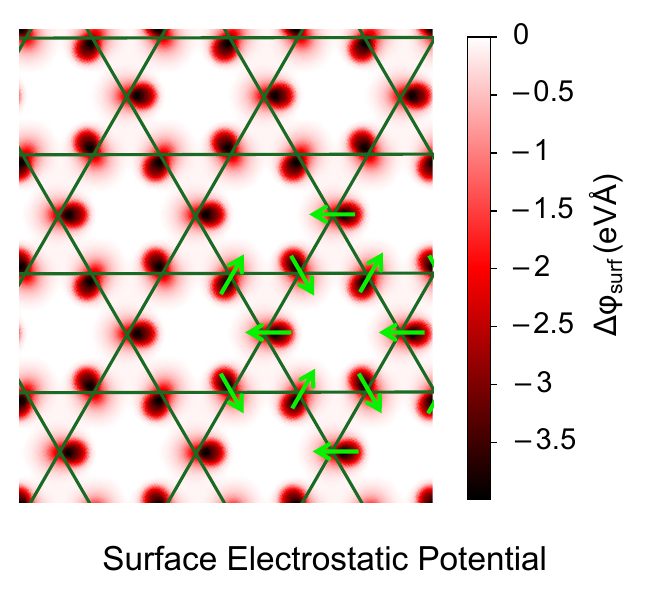}
    \caption{\small\textbf{In-plane electrostatic potential of the top H$_2$O overlayer.}
    Dark regions show negative potential of the H$_2$O-passivated surface relative to the pristine slab, both referenced to their respective vacuum level.
    The potential exhibits pronounced spatial modulations reflecting the positions and orientations of the water molecules and their dipole moments (bright green arrows), forming a frustrated electrostatic landscape on the underlying kagome lattice indicated in dark green.
    }
    \label{fig4}
\end{figure}

{Another interesting finding in these systems is related to their surface electrostatics.}
Local variations in potential can modulate the energies and localization of the surface states, including the kagome-derived flat band.
The electrostatic potential difference $\Delta \varphi_\mathrm{surf}$ of the H$_2$O-passivated surface with respect to its host slab, {obtained by integrating along the surface normal over the top overlayer}, is shown in Fig.~\ref{fig4}.
The potential exhibits strong lateral variations corresponding to the positions and orientations of the water molecules, reflecting the local dipole moments of each adsorbed molecule.
These variations create a highly heterogeneous electrostatic landscape across the surface.

The spatial pattern of the potential difference thus suggests a frustrated arrangement of the water dipoles within the kagome lattice.
Neighboring molecules orient their dipoles to minimize local energy, but the geometric constraints of the lattice as well as the locked orientation of the molecules discussed above prevent full cancellation, leading to frustration.
This complex lateral structure complements the charge redistribution observed in Figs.~\ref{fig2}\textbf{b}--\textbf{d}, providing a clear visualization of the surface electrostatics induced by the adsorbed water molecules.

\section{Conclusions}
We have presented a systematic first-principles investigation of water passivation of non-vdW 2D materials.
By screening a broad {range of systems}, we show that interaction with water is energetically favorable for many non-vdW {materials} and leads to dynamically stable passivated structures.
In contrast to commonly assumed dissociative adsorption, for ternary systems, water remains intact and forms ordered molecular overlayers.

A particularly striking outcome is the emergence of kagome-ordered water lattices on selected surfaces.
These molecular arrangements give rise to flat electronic bands {at or} near the valence-band edge that are predominantly of molecular character.
{The flat bands arise from the kagome topology of the H$_2$O lattice rather than from strong hybridization with the {2D host}, making them robust even within the valence band manifold of the host material, while the same locked dipole orientations give rise to an electrostatically frustrated surface potential.}

More broadly, our results highlight molecular water as a ubiquitous and functionally active component at non-vdW 2D surfaces, especially under solution-processed or ambient conditions.
Beyond surface stabilization, self-assembled molecular lattices provide a chemically intuitive route to engineer surface electronic structure without relying on moir\'e superlattices or structural patterning.
These findings thus open perspectives for exploiting molecular passivation as a design principle for electronic and interfacial properties.

\section{Methods}
Initial geometries of passivated {materials} {are} generated from the structures outlined in Refs.~\cite{Friedrich_NanoLett_2022,Barnowsky_AdvElMats_2023} (excluding broken symmetry systems).
OH$^-$ groups {are} added to the exposed surface cations while H$^+$ cations {are} added to the O, S, and Cl anions of the 2D structure.
The atoms {are} placed in the positions of the bonds broken in the bulk structure upon exfoliation of the non-vdW sheet.
For optimal preconditioning, initial bond lengths {are} estimated using tabulated covalent radii~\cite{Cordero_DT_2008} as implemented in the atomic simulation environment (ASE)~\cite{Larsen_JPhysCM_2017}.
This yields a slight bias towards a dissociated water configuration, leading to {recombined} H$_2$O molecules only when this is clearly energetically favored during relaxation.

All density functional theory (DFT) calculations are performed using the Vienna \emph{Ab-Initio} Simulation Package (VASP)~\cite{vasp_prb1993,Bloechl1994a,vasp_JPCM_1994,vasp_prb1996,vasp_cms1996} with the PBE($+U$) functional~\cite{PBE,Anisimov_Mott_insulators_PRB1991,LDAU1,LDAU2}, following the AFLOW standard~\cite{Calderon_cms_2015} to ensure compatibility with the AFLOW database~\cite{Esters_CMS_2023}.
AFLOW uses an automatic algorithm to generate $\vec k$-point grids for bulk materials~\cite{Oses_CMS_2023, Divilov_HighEntAlloyMat_2025}.
For the hexagonal 2D cells in this work, a modified version {is} used, producing uniform in-plane $n \times n \times 1$ grids, depending on the cell dimensions and number of atoms.
During structure optimization, the out-of-plane lattice vector {is} kept fixed, ensuring at least \SI{20}{\angstrom} interlayer separation.
For magnetic systems the preferred magnetic order {is} determined by explicitly evaluating the energies of all possible magnetic configurations within the primitive unit cell~\cite{Friedrich_CCE_2019}.

For the systems with negative $\Delta E_\mathrm{pas}$, MD simulations {are} performed for \SI{5}{\pico\second} using $2 \times 2$ in-plane supercells containing 112 atoms.
An Andersen thermostat combined with $\Gamma$-point $\vec k$-space sampling {is} used with the time step set to \SI{0.5}{\femto\second}.
As in the structural optimizations, all degrees of freedom {are} allowed to evolve freely during the simulation except for the length and orientation of the out-of-plane lattice vector.
{
Systems {are} considered dynamically stable if the root mean squared (RMS) deviation from the energy of the first MD step is less than \SI{10}{\milli\eV\per atom} over duration of the simulation.
}

{
The electronic structure of the isolated H$_2$O overlayer shown in Fig.~\ref{fig3}\textbf{d} {is} evaluated by retaining the atomic coordinates of the water molecules as obtained from the relaxed passivated geometry, while removing {the underlying 2D material}.
Different intermolecular distances $d$ {are} realized by uniformly scaling the in-plane lattice constant, thereby modifying the separation between neighboring molecular centers, however preserving all intramolecular bond lengths and angles.
These calculations {are} performed using a $10 \times 10 \times 1$ in-plane $\vec k$-point mesh.
}

The projected electrostatic potential shown in Fig.~\ref{fig4} is defined as
\begin{equation}
    \Delta \varphi_\mathrm{surf}(x,y) = \int_{z_1}^{z_2} \varphi_\mathrm{passivated}(\vec r)-\varphi^\mathrm{static}_\mathrm{pristine}(\vec r) \,\dd z, \label{eq:dphi}
\end{equation}
where the integration boundaries $z_{1/2}$ are chosen accordingly to include the electron configuration of only the top overlayer of water molecules.
{
The electrostatic potentials $\varphi$ are the Hartree potentials as determined from the DFT charge density, referenced to their respective vacuum level.
}

{
To assess the influence of dispersive interactions, which are not included in the PBE($+U$) functional used throughout, we additionally {relax} the water-passivated KSbO$_3$ structure with a D3 dispersion correction~\cite{Grimme_JCP_2010}.
Both the resulting structure and the electronic band structure are essentially unchanged compared to the dispersion-uncorrected results, dispersion corrections {are} therefore not included in the production calculations.
}

In addition to standard vacuum calculations, we {employ} VASPsol++~\cite{Islam_JCP_2023} (ISOL = 2) to compute solvation-corrected passivation energies and band structures for the 11 compounds at the end of the employed high-throughput filtering scheme.
The implicit solvent calculations account for the electrostatic screening and stabilization provided by a surrounding aqueous environment.
We {find} a mean absolute error (MAE) of the computed passivation energies versus those determined including solvation of \SI{47.5}{\milli\eV \per H_2O} -- an order of magnitude lower than typical passivation energies.
Hence, the solvation model {does not lead} to major changes in the thermodynamics and structural preferences of the adsorbed molecules, indicating that the {main} trends are already captured in the vacuum calculations.
The solvation corrected passivation energies and band structures can be found in Sec.~II in the SI.

{
The authors acknowledge the use of the large language models (LLMs) ChatGPT and Claude to assist with the refinement of the manuscript text. All scientific content, analysis, and conclusions are the responsibility of the authors.
}

\section*{Data Availability}
The primary research data of this study is made publicly available via the Rossendorf Data Repository (RODARE) at \verb|https://doi.org/10.14278/rodare.4441|~\cite{Barnowsky_Rodare_2026}.

\begin{acknowledgments}
The authors thank Chen-Chen Er, {Balaram Thakur, Thomas Brumme, and Hongde Yu} for fruitful discussions as well as Knut Vietze and Lyuben Zhechkov for technical support.
R.~F. acknowledges funding by the German Research Foundation (DFG), project FR 4545/2-1, collaborative research center SFB 1415, Project C08, Project ID 417590517, and for the ``Autonomous Materials Thermodynamics'' (AutoMaT) project by Technische Universit\"at Dresden and HZDR within the DRESDEN-concept alliance.
M.~C. thanks Helmholtz-Zentrum Dresden-Rossendorf (HZDR) for support through the Summer Student Program.
The authors thank the HZDR Computing Center, HLRS Stuttgart, and the Paderborn Center for Parallel Computing PC$^2$ for generous grants of CPU time.
\end{acknowledgments}

\section*{Author Contributions}
R.~F. initiated and supervised the project.
T.~B. {coordinated} the computational investigation, analyzed, and integrated the results.
T.~B. and R.~F. developed the physical interpretation.
M.~C. performed passivation energy calculations.
A.~N. performed solvation model calculations.
T.~B. wrote the initial draft of the manuscript, which was refined by R.~F.
All authors contributed to the discussion and writing of the final manuscript.

\section*{Competing Interests}
{The authors declare no competing interests.}

\newcommand{\Ozolins}{Ozoli{\c{n}}{\v{s}}}

\clearpage
\FloatBarrier
\onecolumngrid
\begin{center}
  \Large
  \textbf{
  Supplementary Information
  }
\end{center}

\setcounter{figure}{0}
\setcounter{table}{0}
\setcounter{section}{0}
\renewcommand{\thesection}{S\Roman{section}}
\renewcommand{\thetable}{S\Roman{table}}
\renewcommand{\thefigure}{S\arabic{figure}}

\section{Band Structures}

\newcommand{\X}{0.9}

\begin{figure}[ht]
  \centering
  \includegraphics[width=\linewidth]{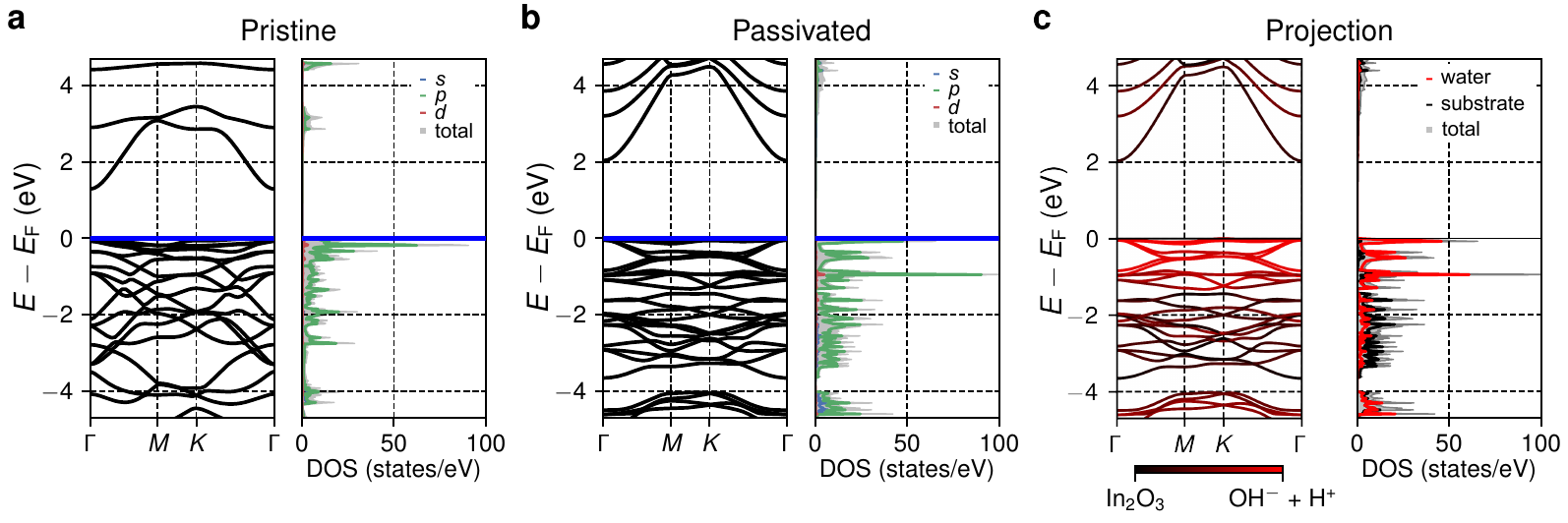}
  \caption{
    Band structure and DOS of pristine (\textbf{a}) and hydrated (\textbf{b}) In$_2$O$_3$.
    For the passivated system, the projection onto the dissociated water states (OH$^-$ + H$^+$) is shown in (\textbf{c}).
  }
  \label{fig:SI-1}
\end{figure}

\begin{figure}[ht]
  \centering
  \includegraphics[width=\linewidth]{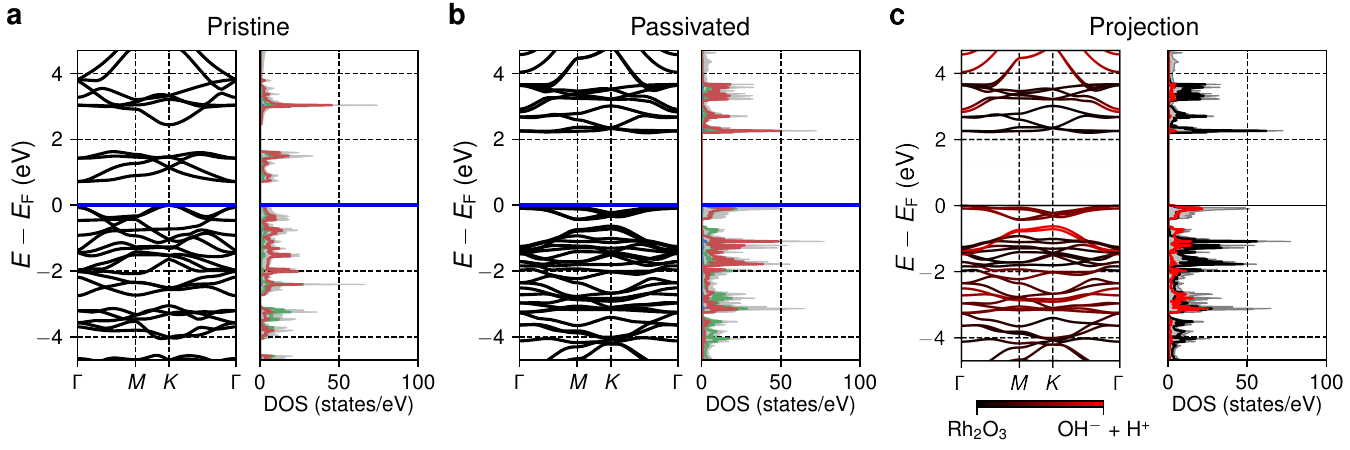}
  \caption{
    Band structure and DOS of pristine (\textbf{a}) and hydrated (\textbf{b}) Rh$_2$O$_3$.
    For the passivated system, the projection onto the dissociated water states (OH$^-$ + H$^+$) is shown in (\textbf{c}).
  }
  \label{fig:SI-2}
\end{figure}

\begin{figure}[ht]
  \centering
  \includegraphics[width=\linewidth]{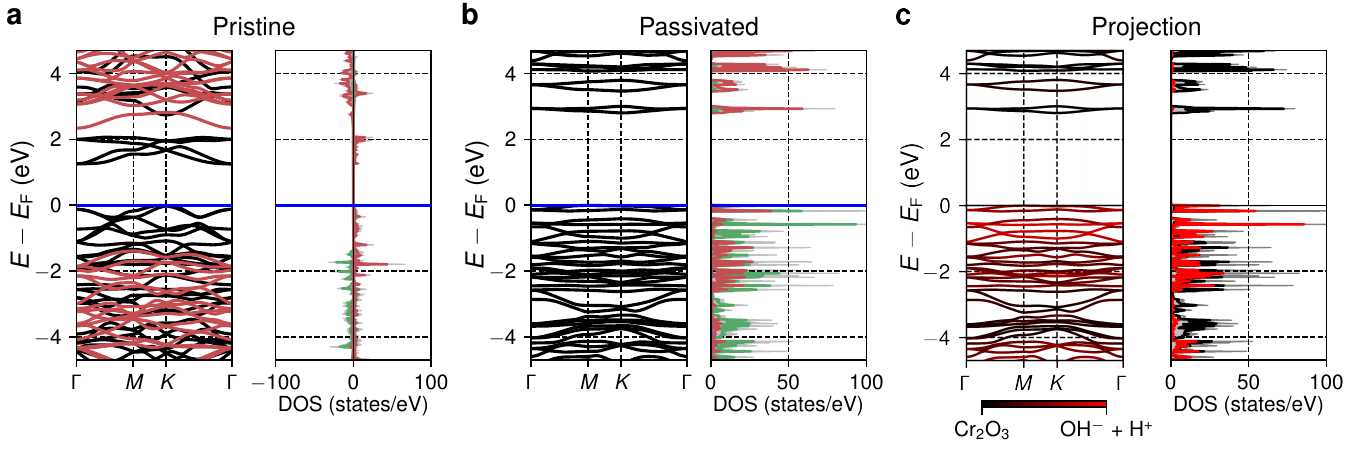}
  \caption{
    Band structure and DOS of pristine (\textbf{a}) and hydrated (\textbf{b}) Cr$_2$O$_3$.
    For the passivated system, the projection onto the dissociated water states (OH$^-$ + H$^+$) is shown in (\textbf{c}).
    Negative/positive DOS correspond to spin-up/down states respectively.
  }
  \label{fig:SI-3}
\end{figure}

\begin{figure}[ht]
  \centering
  \includegraphics[width=\linewidth]{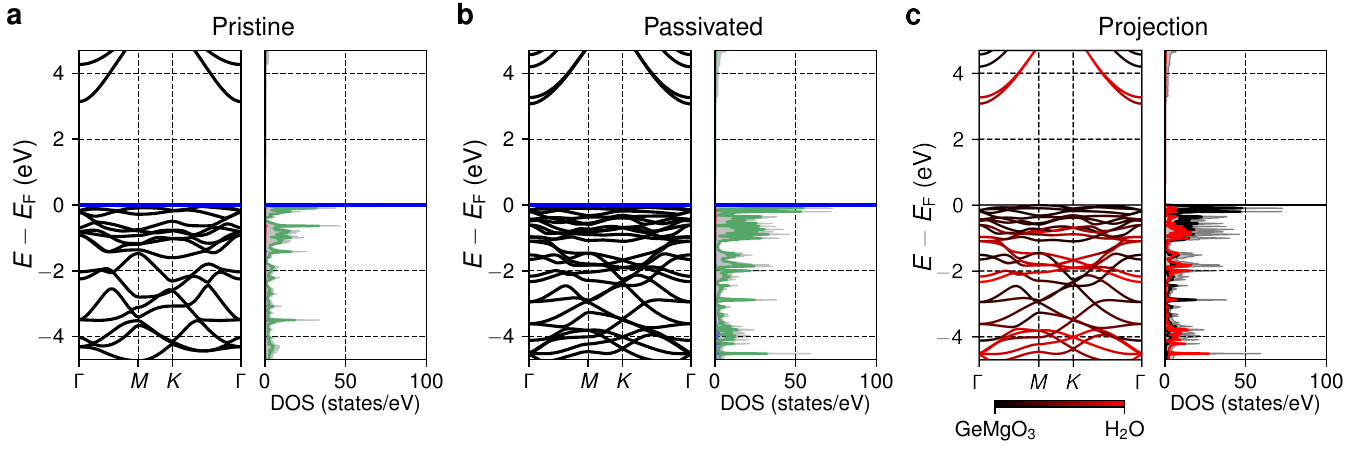}
  \caption{
    Band structure and DOS of pristine (\textbf{a}) and hydrated (\textbf{b}) GeMgO$_3$.
    For the passivated system, the projection onto the water overlayer states (H$_2$O) is shown in (\textbf{c}).
  }
  \label{fig:SI-4}
\end{figure}

\begin{figure}[ht]
  \centering
  \includegraphics[width=\linewidth]{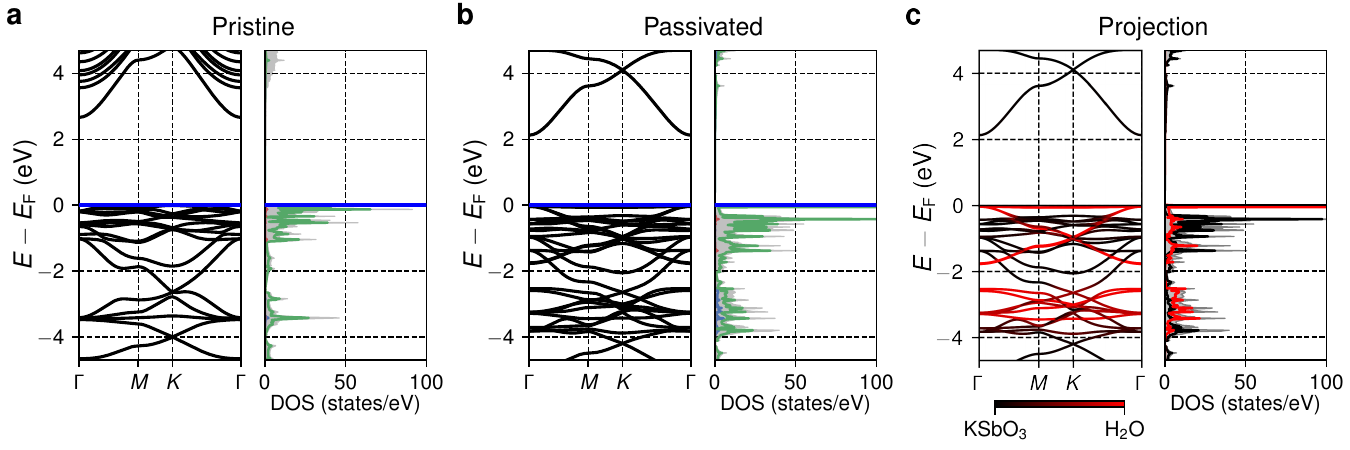}
  \caption{
    Band structure and DOS of pristine (\textbf{a}) and hydrated (\textbf{b}) KSbO$_3$.
    For the passivated system, the projection onto the water overlayer states (H$_2$O) is shown in (\textbf{c}).
  }
  \label{fig:SI-5}
\end{figure}

\begin{figure}[ht]
  \centering
  \includegraphics[width=\linewidth]{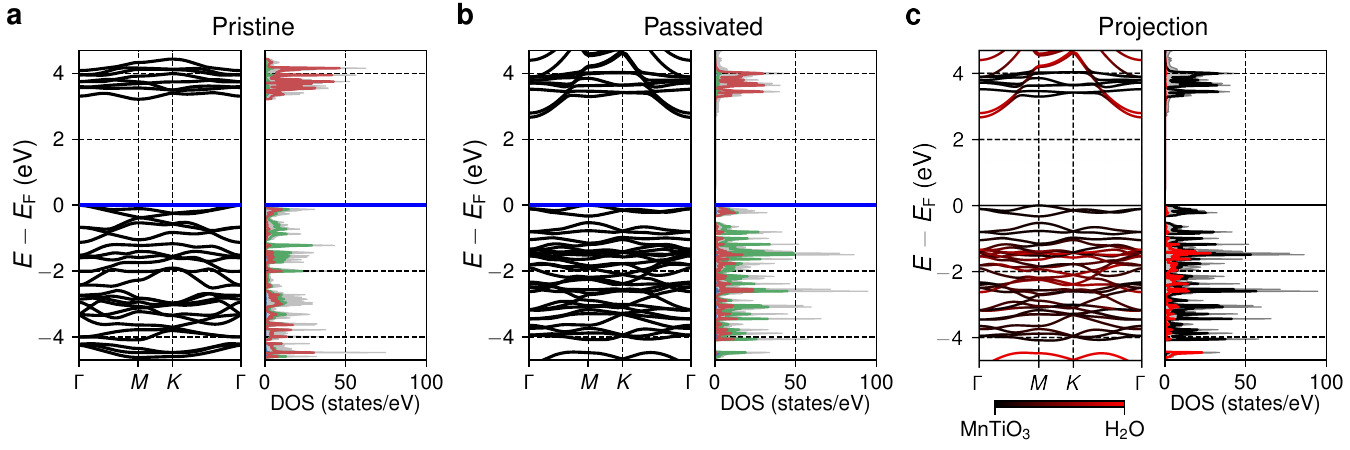}
  \caption{
    Band structure and DOS of pristine (\textbf{a}) and hydrated (\textbf{b}) MnTiO$_3$.
    For the passivated system, the projection onto the water overlayer states (H$_2$O) is shown in (\textbf{c}).
  }
  \label{fig:SI-6}
\end{figure}

\begin{figure}[ht]
  \centering
  \includegraphics[width=\linewidth]{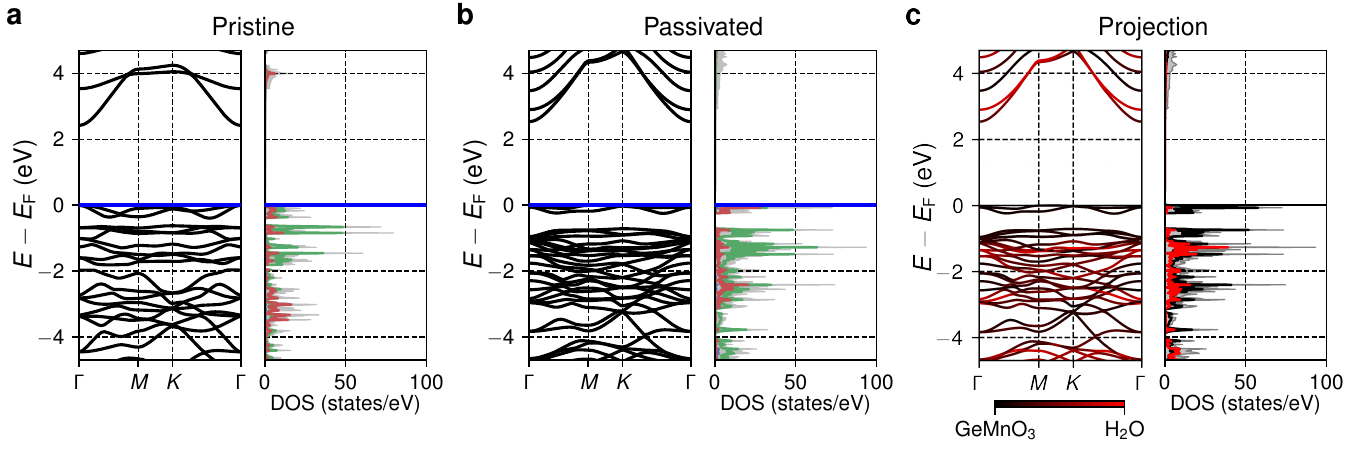}
  \caption{
    Band structure and DOS of pristine (\textbf{a}) and hydrated (\textbf{b}) GeMnO$_3$.
    For the passivated system, the projection onto the water overlayer states (H$_2$O) is shown in (\textbf{c}).
  }
  \label{fig:SI-7}
\end{figure}

\begin{figure}[ht]
  \centering
  \includegraphics[width=\linewidth]{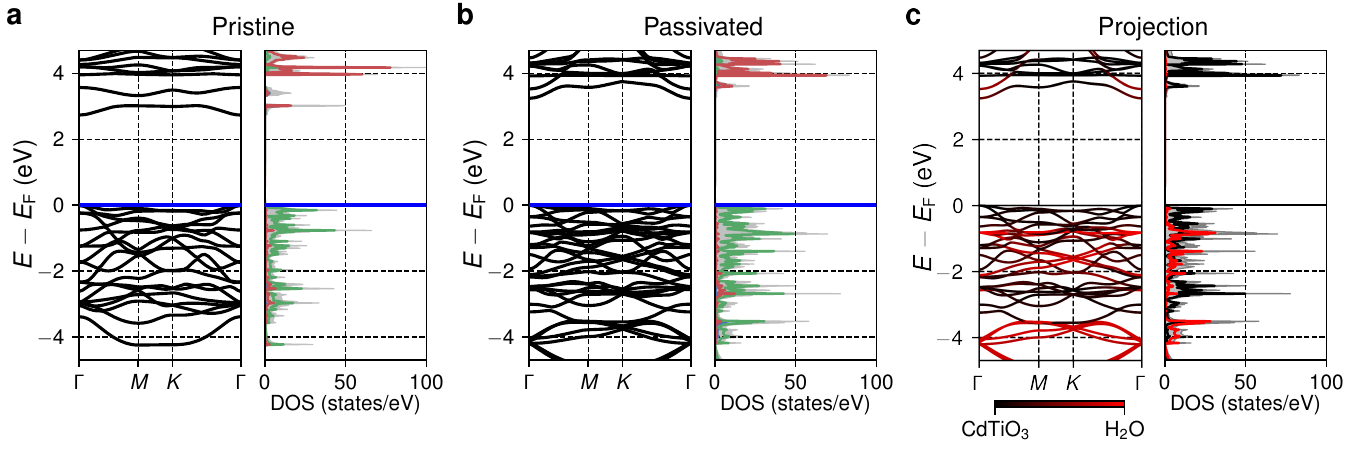}
  \caption{
    Band structure and DOS of pristine (\textbf{a}) and hydrated (\textbf{b}) CdTiO$_3$.
    For the passivated system, the projection onto the water overlayer states (H$_2$O) is shown in (\textbf{c}).
  }
  \label{fig:SI-8}
\end{figure}

\begin{figure}[ht]
  \centering
  \includegraphics[width=\linewidth]{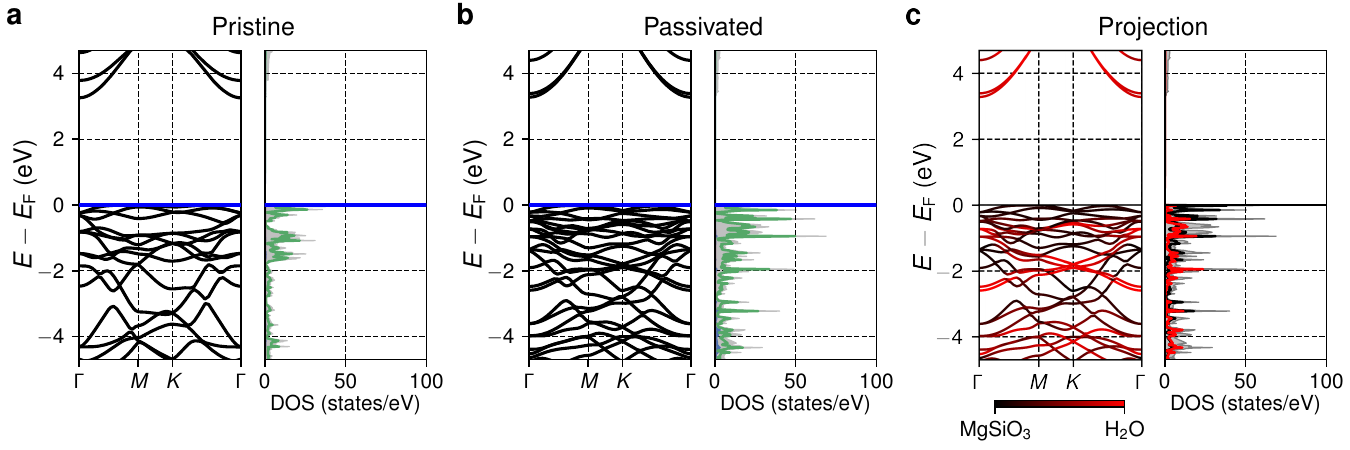}
  \caption{
    Band structure and DOS of pristine (\textbf{a}) and hydrated (\textbf{b}) MgSiO$_3$.
    For the passivated system, the projection onto the water overlayer states (H$_2$O) is shown in (\textbf{c}).
  }
  \label{fig:SI-9}
\end{figure}

\begin{figure}[ht]
  \centering
  \includegraphics[width=\linewidth]{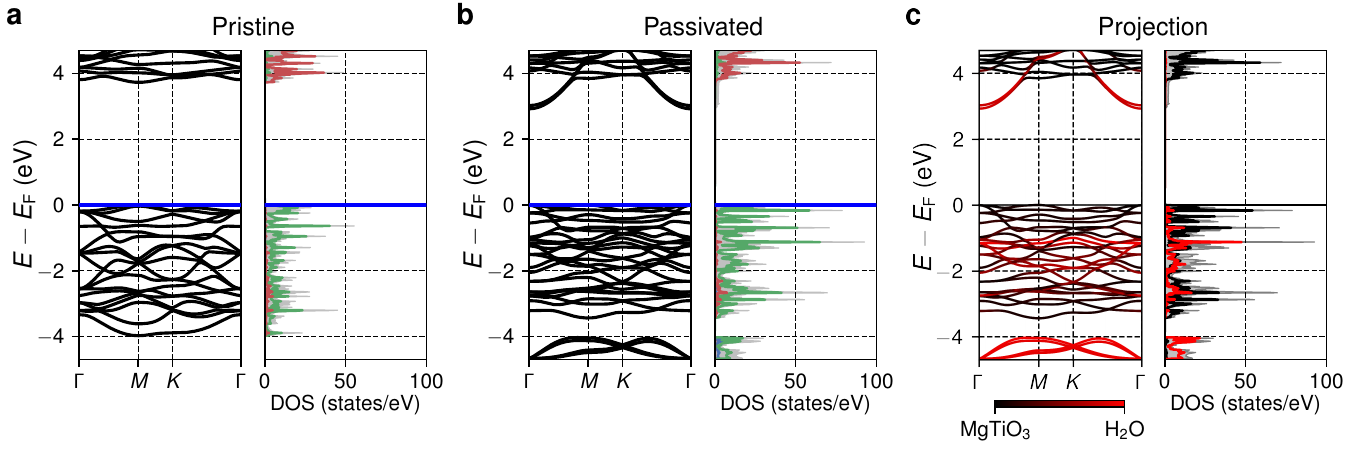}
  \caption{
    Band structure and DOS of pristine (\textbf{a}) and hydrated (\textbf{b}) MgTiO$_3$.
    For the passivated system, the projection onto the water overlayer states (H$_2$O) is shown in (\textbf{c}).
  }
  \label{fig:SI-10}
\end{figure}

\begin{figure}[ht]
  \centering
  \includegraphics[width=\linewidth]{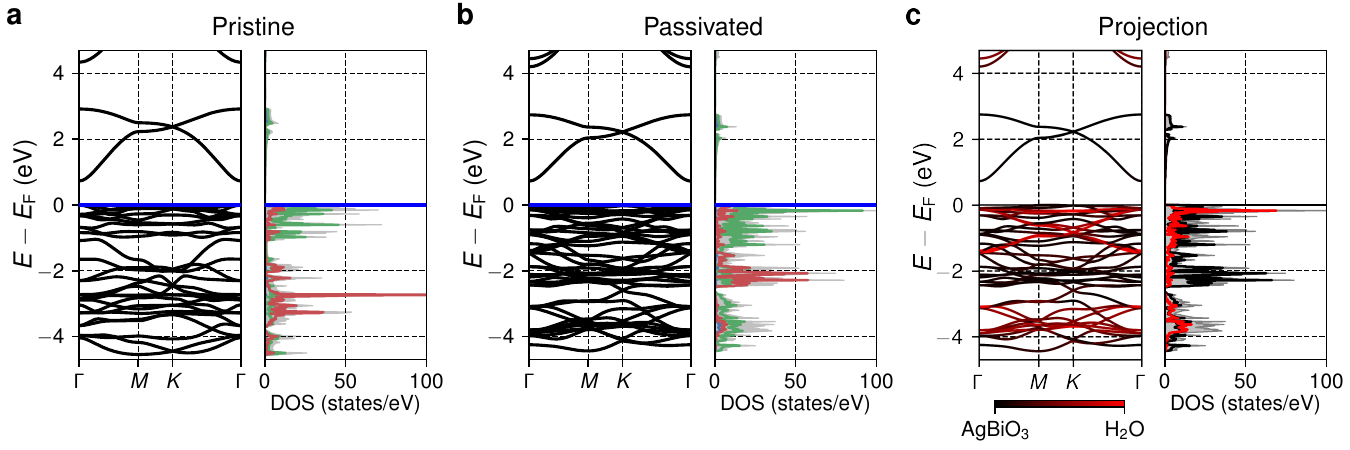}
  \caption{
    Band structure and DOS of pristine (\textbf{a}) and hydrated (\textbf{b}) AgBiO$_3$.
    For the passivated system, the projection onto the water overlayer states (H$_2$O) is shown in (\textbf{c}).
  }
  \label{fig:SI-11}
\end{figure}

\FloatBarrier
\section{Solvation Effects}

\subsection{Passivation Energies}

\begin{figure}[ht]
  \centering
  \includegraphics[width=\linewidth]{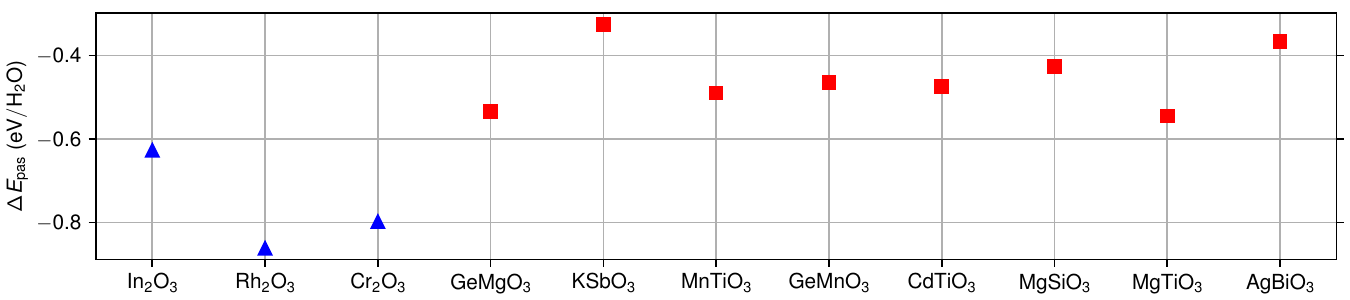}
  \caption{
    Solvation corrected passivation energies for the final set of thermodynamically favored and dynamically stable of non-vdW 2D materials upon hydration.
    Systems for which water dissociates upon binding are marked with blue triangles, while systems where water molecules remain intact are shown as red squares.
  }
  \label{fig:SI-12}
\end{figure}

\subsection{Band Structures}

\begin{figure}[ht]
  \centering
  \includegraphics[width=\linewidth]{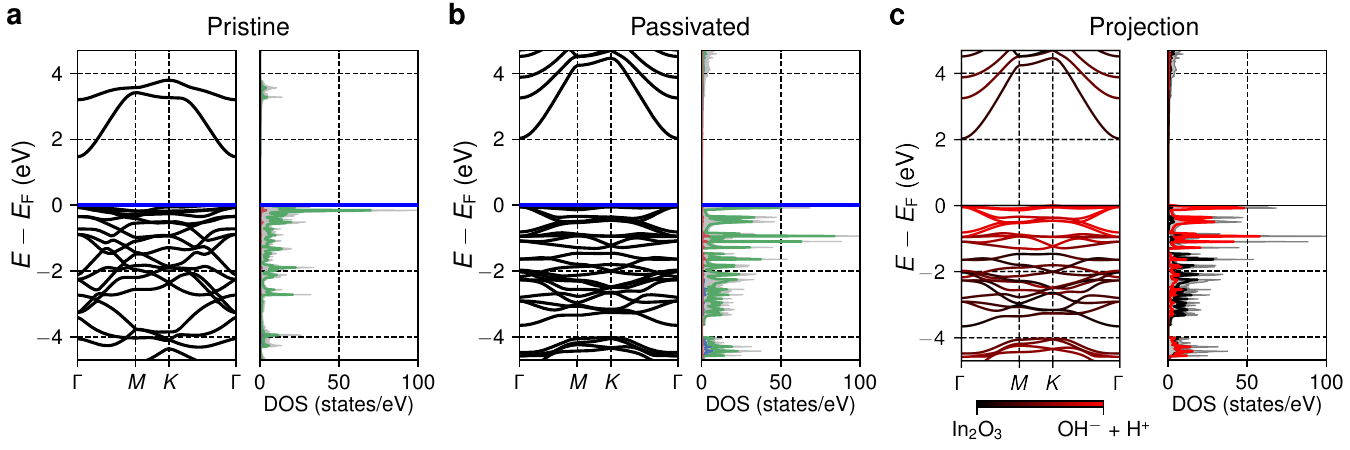}
  \caption{
    Solvation corrected band structure and DOS of pristine (\textbf{a}) and hydrated (\textbf{b}) In$_2$O$_3$.
    For the passivated system, the projection onto the dissociated water states (OH$^-$ + H$^+$) is shown in (\textbf{c}).
  }
  \label{fig:SI-13}
\end{figure}

\begin{figure}[ht]
  \centering
  \includegraphics[width=\linewidth]{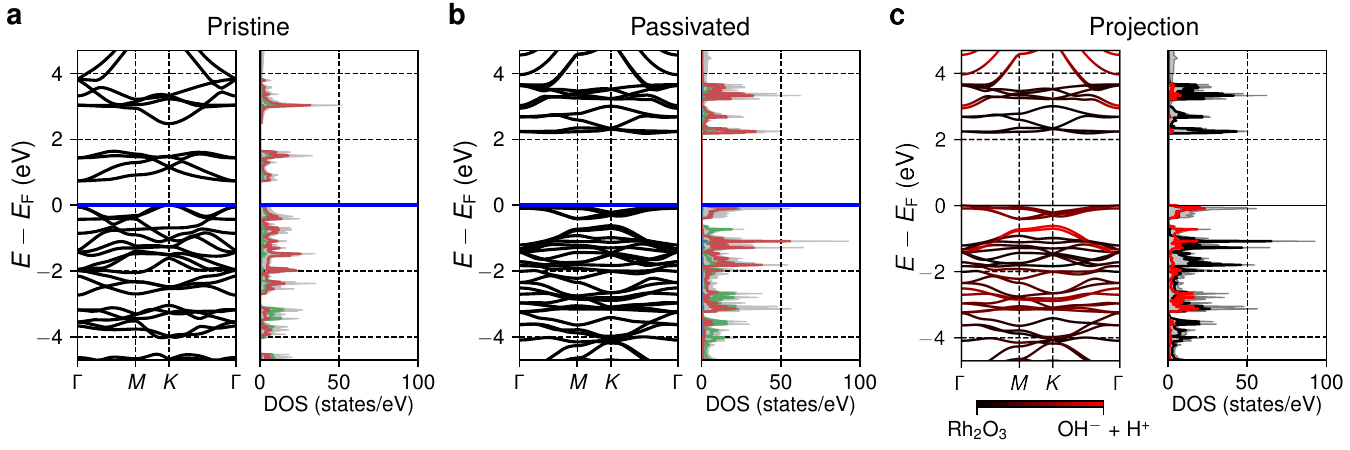}
  \caption{
    Solvation corrected band structure and DOS of pristine (\textbf{a}) and hydrated (\textbf{b}) Rh$_2$O$_3$.
    For the passivated system, the projection onto the dissociated water states (OH$^-$ + H$^+$) is shown in (\textbf{c}).
  }
  \label{fig:SI-14}
\end{figure}

\begin{figure}[ht]
  \centering
  \includegraphics[width=\linewidth]{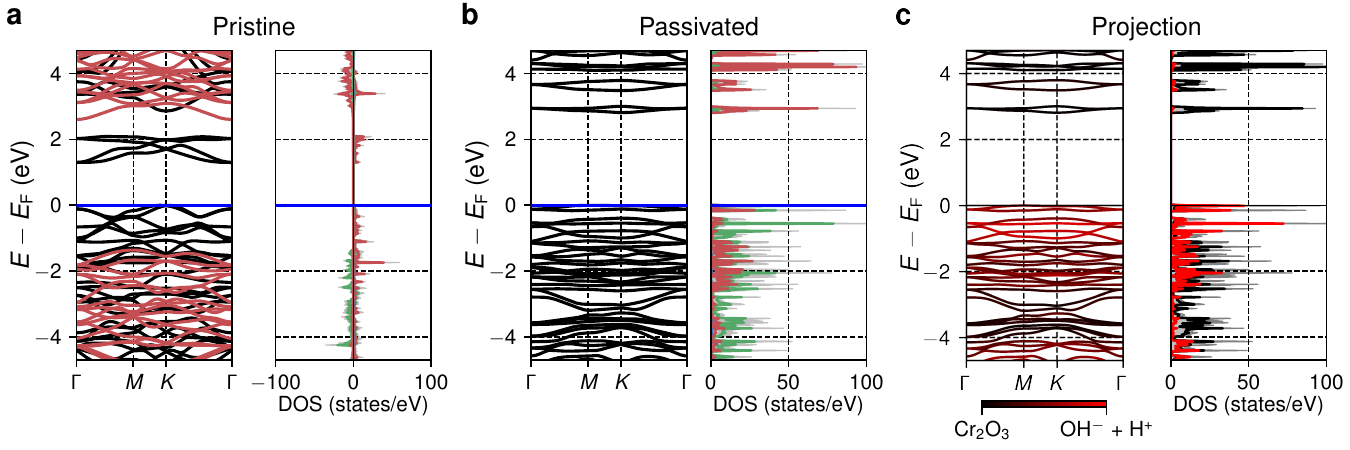}
  \caption{
    Solvation corrected band structure and DOS of pristine (\textbf{a}) and hydrated (\textbf{b}) Cr$_2$O$_3$.
    For the passivated system, the projection onto the dissociated water states (OH$^-$ + H$^+$) is shown in (\textbf{c}).
    Negative/positive DOS correspond to spin-up/down states respectively.
  }
  \label{fig:SI-15}
\end{figure}

\begin{figure}[ht]
  \centering
  \includegraphics[width=\linewidth]{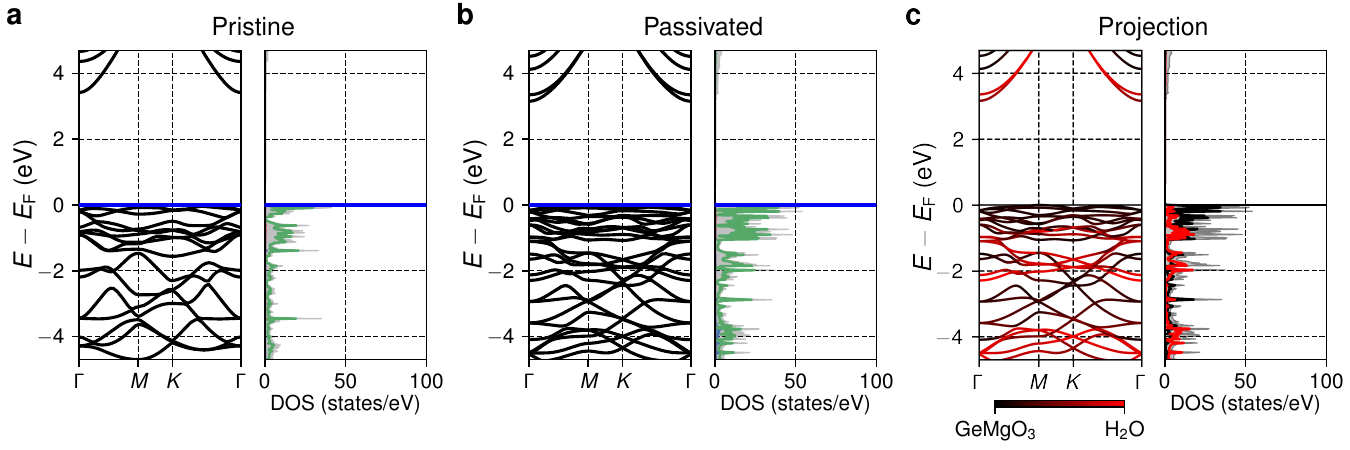}
  \caption{
    Solvation corrected band structure and DOS of pristine (\textbf{a}) and hydrated (\textbf{b}) GeMgO$_3$.
    For the passivated system, the projection onto the water overlayer states (H$_2$O) is shown in (\textbf{c}).
  }
  \label{fig:SI-16}
\end{figure}

\begin{figure}[ht]
  \centering
  \includegraphics[width=\linewidth]{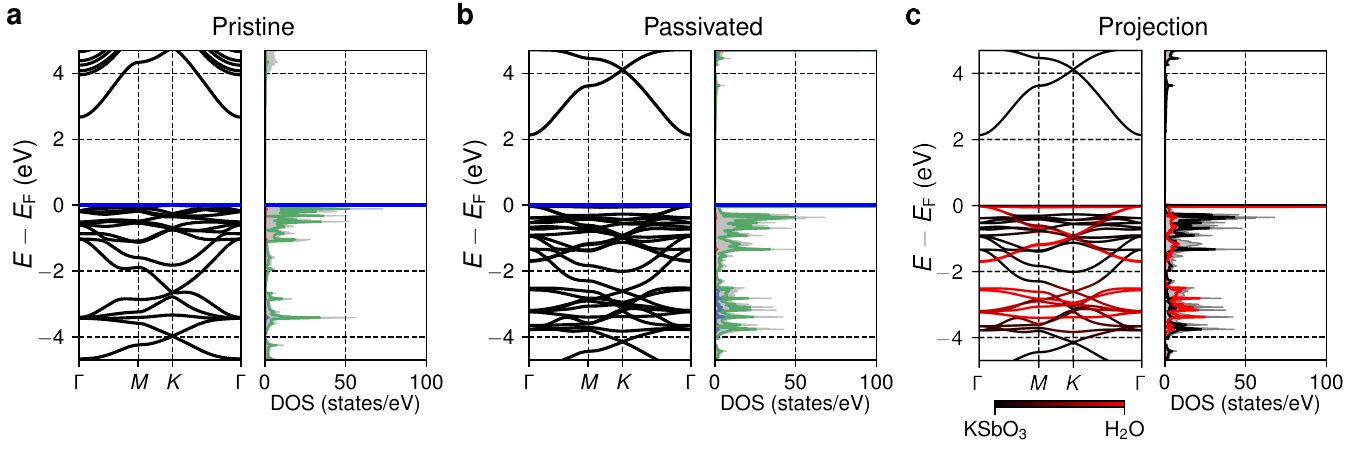}
  \caption{
    Solvation corrected band structure and DOS of pristine (\textbf{a}) and hydrated (\textbf{b}) KSbO$_3$.
    For the passivated system, the projection onto the water overlayer states (H$_2$O) is shown in (\textbf{c}).
  }
  \label{fig:SI-17}
\end{figure}

\begin{figure}[ht]
  \centering
  \includegraphics[width=\linewidth]{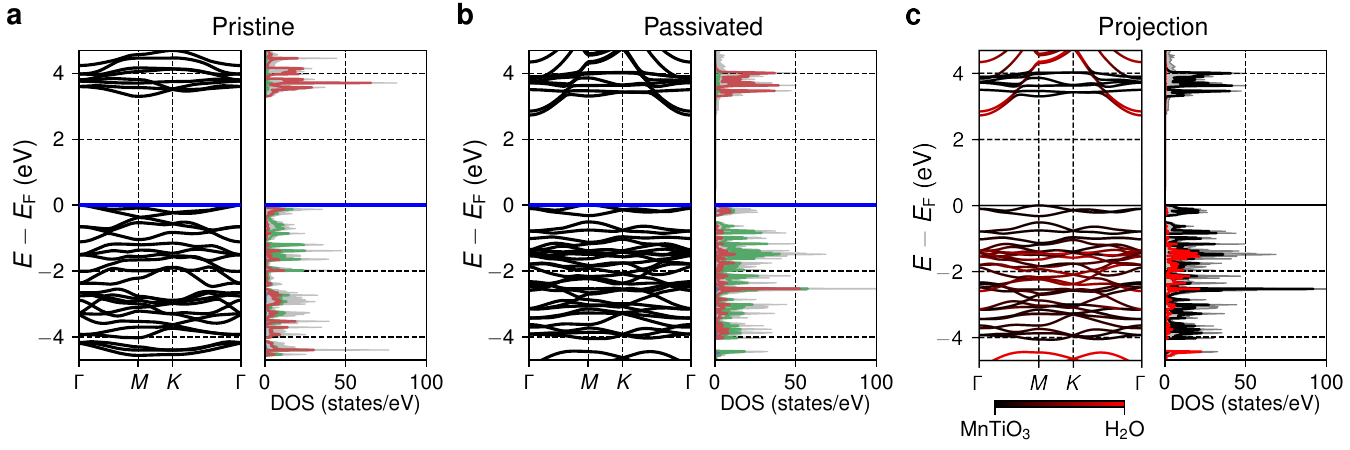}
  \caption{
    Solvation corrected band structure and DOS of pristine (\textbf{a}) and hydrated (\textbf{b}) MnTiO$_3$.
    For the passivated system, the projection onto the water overlayer states (H$_2$O) is shown in (\textbf{c}).
  }
  \label{fig:SI-18}
\end{figure}

\begin{figure}[ht]
  \centering
  \includegraphics[width=\linewidth]{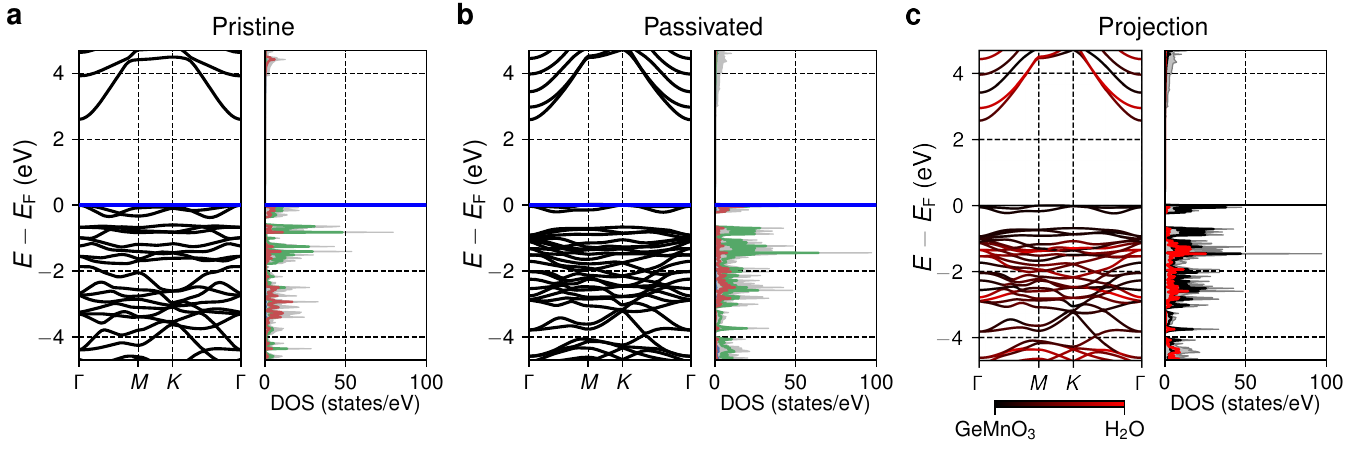}
  \caption{
    Solvation corrected band structure and DOS of pristine (\textbf{a}) and hydrated (\textbf{b}) GeMnO$_3$.
    For the passivated system, the projection onto the water overlayer states (H$_2$O) is shown in (\textbf{c}).
  }
  \label{fig:SI-19}
\end{figure}

\begin{figure}[ht]
  \centering
  \includegraphics[width=\linewidth]{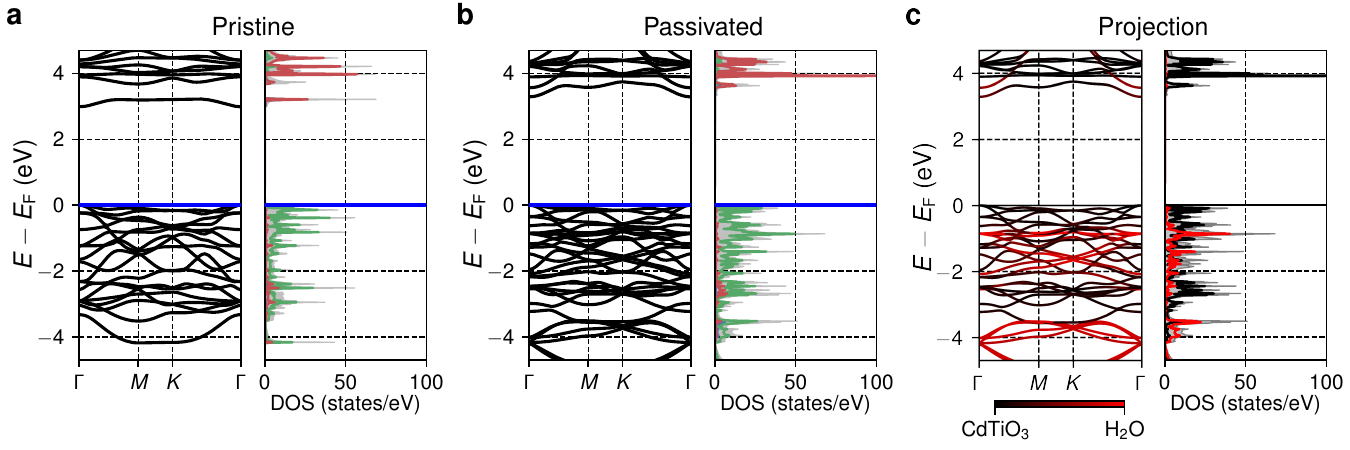}
  \caption{
    Solvation corrected band structure and DOS of pristine (\textbf{a}) and hydrated (\textbf{b}) CdTiO$_3$.
    For the passivated system, the projection onto the water overlayer states (H$_2$O) is shown in (\textbf{c}).
  }
  \label{fig:SI-20}
\end{figure}

\begin{figure}[ht]
  \centering
  \includegraphics[width=\linewidth]{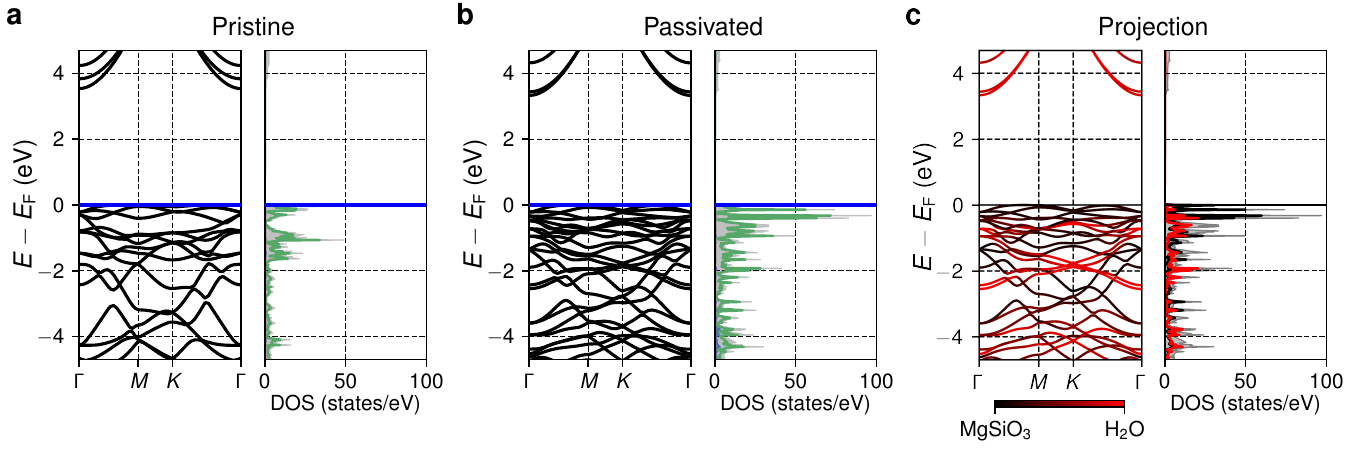}
  \caption{
    Solvation corrected band structure and DOS of pristine (\textbf{a}) and hydrated (\textbf{b}) MgSiO$_3$.
    For the passivated system, the projection onto the water overlayer states (H$_2$O) is shown in (\textbf{c}).
  }
  \label{fig:SI-21}
\end{figure}

\begin{figure}[ht]
  \centering
  \includegraphics[width=\linewidth]{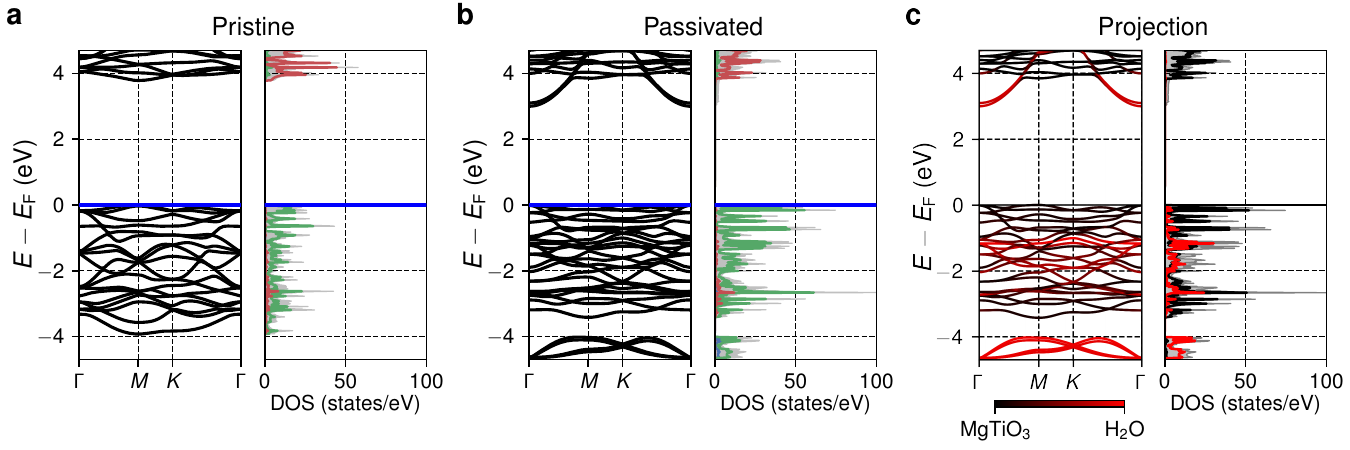}
  \caption{
    Solvation corrected band structure and DOS of pristine (\textbf{a}) and hydrated (\textbf{b}) MgTiO$_3$.
    For the passivated system, the projection onto the water overlayer states (H$_2$O) is shown in (\textbf{c}).
  }
  \label{fig:SI-22}
\end{figure}

\begin{figure}[ht]
  \centering
  \includegraphics[width=\linewidth]{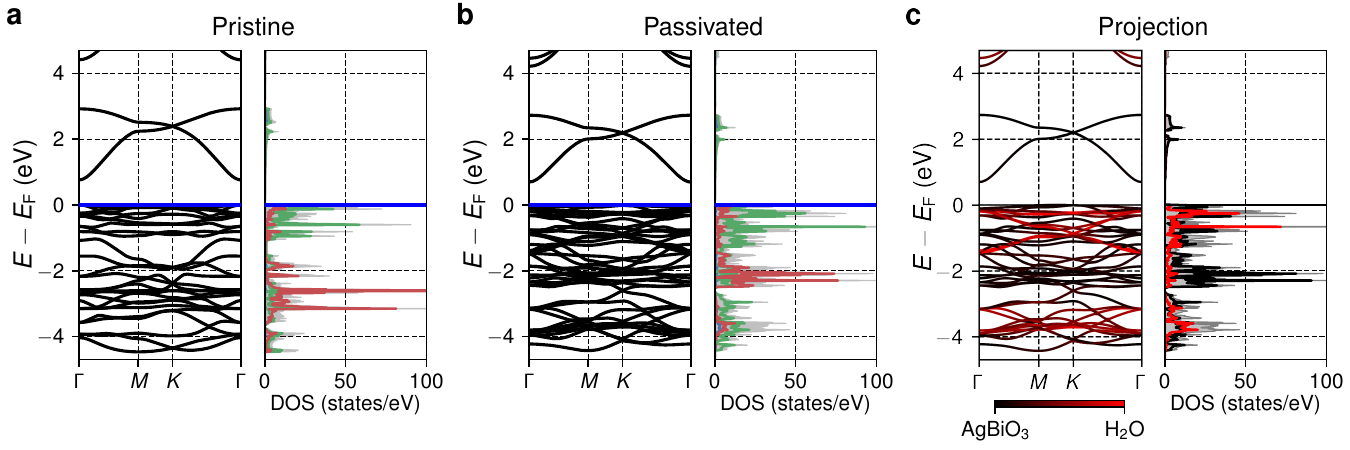}
  \caption{
    Solvation corrected band structure and DOS of pristine (\textbf{a}) and hydrated (\textbf{b}) AgBiO$_3$.
    For the passivated system, the projection onto the water overlayer states (H$_2$O) is shown in (\textbf{c}).
  }
  \label{fig:SI-23}
\end{figure}

\end{document}